\documentclass{ws-procs975x65}
\begin{document}
\title{GEODESIC FLOWS AND THEIR DEFORMATIONS IN BERTRAND SPACETIMES}
\author{PRASHANT KUMAR, KAUSHIK BHATTACHARYA, TAPOBRATA SARKAR}
\address{
\normalsize
Department of Physics, Indian Institute of Technology, Kanpur,\\
\normalsize
Kanpur 208016, India,\\
kaushikb@iitk.ac.in}
\begin{abstract}
In this article we will discuss some features of a particular
spacetime called Bertrand space-time of Type II (BST-II).  The talk
discusses about the energy conditions and the ESR parameters in
this spacetime. 
\end{abstract}
\bodymatter
\bigskip
\section{The energy conditions in BST-II spacetime}
There has been an attempt \cite{Kumar} to find out the properties of potentials
which can produce closed, stable orbits using special relativistic
techniques.  In general relativistic parlance the particular spacetime
\begin{eqnarray}
ds^2= -\frac{dt^2}{G +\sqrt{r^{-2}+K}} + \frac{dr^2}{\beta^2(1+Kr^2)}
+r^2(d\theta^2 + \sin^2 \theta\,d\phi^2)\,,  
\label{type2}
\end{eqnarray}
called Bertrand spacetime\cite{perlick} of Type II (BST-II) has the 
property of admitting periodic closed orbits.  The
parameters $D$, $G$ and $K$ are real, and $\beta$ must be a positive
rational number.  The specialty of BST-II proposed by Perlick, in
Ref.~[\citen{perlick, enciso}], is that this kind of spacetime admits
closed, stable and periodic orbits from any point in the
manifold. Unlike Schwarszchild solution BST-II is not a vacuum
solution of Einstein's equation. The work presented in this talk rests
heavily on Ref.~\citen{Kumar:2012zt}.

The energy density, in BST-II, turns out to be
\begin{equation}
\rho = \frac{1-\beta^2 \left(3 K r^2+1\right)}{r^2}
\label{rhokneq0}
\end{equation}
For $r \gg 1$, this implies that the energy density is negative for
positive values of $K$. The situation might be remedied by choosing a
negative value of $K$. One can note that this necessitates, from
Eq.~(\ref{type2}) that for $K = -\kappa$ (where $\kappa$ is a positive
real number), $r < 1/\sqrt{\kappa}$. We can thus choose $\kappa \ll 1$
so that the positivity of the energy density of space-time of
Eq.~(\ref{type2}) is guaranteed for a large range of $r$. The analysis
of the weak energy condition (WEC) shows that a given choice of
$\kappa$ (in accordance with the discussion above), the WEC is always
satisfied for $r < 1/\sqrt{\kappa}$ when $G$ is a small positive real number.

Before we end this section, we will briefly comment on the strong
energy condition (SEC): $\rho + \sum_i p_i \geq 0,~~\rho + p_i \geq
0$. We find that for the metric of Eq..(\ref{type2}),
\begin{equation}
\rho + \sum_ip_i = \frac{3 \beta^2}{2 r^2 \left(G r+\sqrt{K r^2+1}\right)^2}
\end{equation}
so that the SEC is satisfied whenever $r < 1/\sqrt{\kappa}$, for
positive values of $G$. 
\section{Geodesics flows in BST-II spacetime}
Calling the deformation (between two points on two geodesics) vector
$\xi^\mu$ one can write $\dot{\xi^\mu}=B^\mu_{\,\,\,\nu}\xi^\nu$,
where $B^\mu_{\,\,\,\nu}$ is a second rank tensor. From the last equation  
one can write:
\begin{eqnarray}
\ddot{\xi^\mu}=(\dot{B}^\mu_{\,\,\,\nu} + B^\mu_{\,\,\,\tau}
B^\tau_{\,\,\,\nu})\xi^\nu\,.
\label{xiddot}
\end{eqnarray}
In $n$ space-time dimensions, the general form of the second rank
tensor $B_{\mu \nu}$ can be decomposed into irreducible parts as
\cite{poisson}:
\begin{eqnarray}
B_{\mu \nu}=\frac{1}{n-1}\Theta h_{\mu\nu} + \sigma_{\mu\nu} + \omega_{\mu\nu}\,,
\label{bdecomp}
\end{eqnarray}
where $h_{\mu\nu} = g_{\mu\nu} + u_{\mu}u_{\nu}$ for $u_{\mu}$
time-like tangent vector on the geodesic, and $\Theta$ is the
expansion variable, $\sigma_{\mu\nu}$ is associated with shear and
$\omega_{\mu\nu}$ signifies rotation.  One can explicitly write
\begin{eqnarray}
\Theta &=& B^\mu_{\,\,\,\mu}\,,
\label{thetaexp}\\
\sigma_{\mu\nu} &=& \frac12 (B_{\mu \nu} + B_{\nu \mu}) - \frac{1}{n-1}
\Theta h_{\mu\nu}\,,
\label{sigmaexp}\\
\omega_{\mu\nu} &=& \frac12 (B_{\mu \nu} - B_{\nu \mu})\,. 
\end{eqnarray}
\begin{figure}[b!]
\centering
\includegraphics[width=10cm,height=8cm]{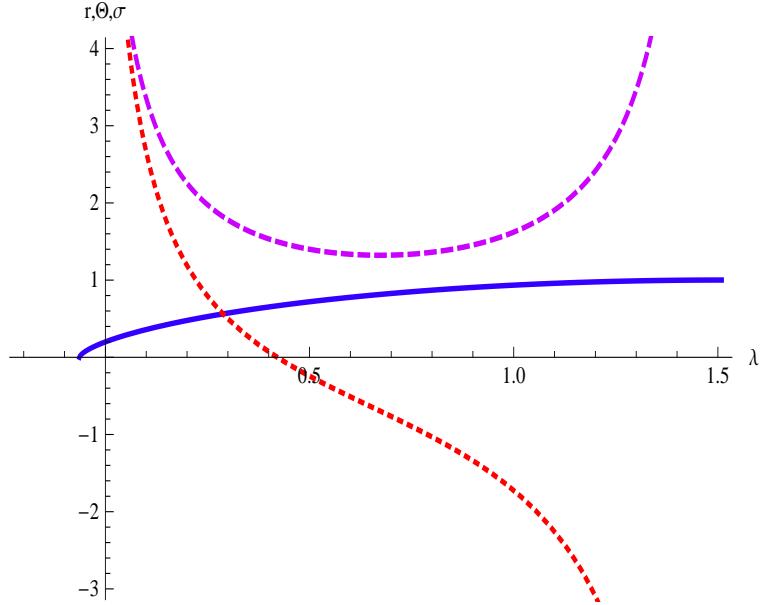}
\caption[]{Numerical solutions for $r$ (solid blue), $\Theta$ (dotted
  red), and $\sigma$ 
(dashed magenta) for radial time-like geodesics in 
BST-II, as a function of $\lambda$ for $K=0$.}
\label{st:f1}
\end{figure}
The ESR variables are generally denoted by $\Theta$, $\sigma\equiv
\sqrt{\sigma^2}$ and $\omega\equiv \sqrt{\omega^2}$. From the
discussions in this section it is seen that
if one knows the form of $u^\mu$ one can calculate $B_{\mu \nu}$, and
hence the ESR parameters as shown in Refs.~[\citen{senguptakar}].

From now we will discuss the geodesics and their properties in the
equatorial plane, i.e. for $\theta=\pi/2$ and $K$ will be assumed to
be a negative real number unless it is set to zero. In general we will
assume $G$ to be a positive real number. The choices are guided by the
energy conditions discussed previously. Using the geodesic equations
and the normalization condition of the 4-velocity $u^\mu \equiv
dx^\mu/d\lambda$ (where $\lambda$ is the affine parameter) the
expression for $u^r$ for the outgoing timelike radial geodesics comes
out as:
\begin{eqnarray}
u^r=
\beta\sqrt{(1+Kr^2)\left[C^2\left(\sqrt{K+r^{-2}}+G\right)-1\right]}\,.
\label{ur}
\end{eqnarray}
In the above expression $C$ is an integration constant appearing in
the geodesic equations. Assuming $Kr^2 + 1 >0$ this implies that there
is a turning point of the outgoing radial time-like geodesics, for
$C^2\left(\sqrt{K + r^{-2}} + G\right) = 1$.  If $K=0$ then the
$\Theta$ parameter for the radial outgoing timelike geodesics looks
like:
\begin{equation}
\Theta = \frac{(G r+1) \left(2 C^2-\frac{r (2 G r+3)}{(G
    r+1)^2}\right)}{2 r^2 \sqrt{C^2 \left(G+\frac{1}{r}\right)-1}}\,.
\label{thetak0}
\end{equation}
One can solve Eq.~(\ref{ur}) in terms of $\lambda$ and then plug in
this solution in the above equation and get $\Theta$ as a function of
the affine parameter. The functional dependence of $r$, $\Theta$ and
$\sigma$ on $\lambda$ is shown in Fig.~\ref{st:f1}.  To draw the above
plot we have taken $G=10^{-3}$, $C=1$, and the lower and upper limits
of the affine parameter have been set to $-0.2$ and $1.5$. The upper
limit of $\lambda$ is chosen so that $r$ varies from zero to the
turning point of $u^r$, which can be seen to be $r \sim 1$ in this
case.  From the figure we see that ${d\theta}/{d\lambda}$ is always
negative. As $\omega=0$ (in this case) this fact confirms the focusing
theorem \cite{poisson}.  Also, $\Theta$ diverges at the upper and
lower limits of $r$, signaling a true spacetime singularity at $r=0$
and the turning point.  The $K \neq 0$ case for radial time-like
geodesics follow the same qualitative behavior.

In our work we did not explicitly solve the Raychaudhuri equation in
BST-II but have expressed the ESR parameters in terms of the affine
parameter $\lambda$ by using the geodesic equations. As finally the
ESR parameters are expressed in terms of $\lambda$ our treatment
indirectly gives the solution of the Raychaudhuri equations in BST-II. 

\end{document}